\documentclass[10pt]{article}
\usepackage{latexsym,graphicx, amssymb}
\newcommand{\be}{\begin{equation}}
\newcommand{\ee}{\end{equation}}
\def\n{\noindent}
\catcode `\@=11
\catcode `\@=12
\begin{document}
\begin{center}
\large{\bf{L.R.S. Bianchi type II stiff fluid cosmological model
 with Decaying Vacuum Energy Density $\Lambda$ in general relativity}} \\
\vspace{10mm}
\normalsize{Hassan Amirhashchi}\\
\vspace{5mm} \normalsize{Department of Physics, Islamic Azad University, Mahshahr Branch, Mahshahr, Iran\\
E-mail : h.amirhashchi@mahshahriau.ac.ir}, hashchi@yahoo.com \\
\end{center}
\vspace{0mm}
\begin{abstract}
Locally rotationally symmetric (L.R.S.) Bianchi type II stiff fluid cosmological model is
investigated. To get the deterministic model of the universe, we have assumed a condition $A=B^{m}$ between metric potentials $A,~B$ where $n$ is the constant. It is shown that the vacuum energy density $\Lambda$ is positive and proportional  to $\frac{1}{t^{2}}$. The values of deceleration parameter $q$, matter-energy density $\Omega_{m}$ and dark-energy density $\Omega_{\Lambda}$ are found to be in good agreement with the values obtain from 5-years WMAP observations. the predicted value of the jerk parameter agrees with the SNLS SNIa and X-ray galaxy cluster distance data but does not with the SNIa gold sample data. In general, the model represent accelerating, shearing and non-rotating universe.The physical and geometrical behavior of these models are also discussed.
\end{abstract}
\smallskip
\n Keywords: LRS Bianchi type II models. Cosmological constant. Stiff fluid\\
\n PACS number: 98.80.Cq, 04.20.-q, 04.20.Jb, 98.80.-k
\section{Introduction}
One of the outstanding problems in cosmology is the so called ``cosmological constant" problem. Recent observations of type Ia supernovae (the Supernova Cosmology Project and the High-Z Supernova Team) \cite{ref1}-\cite{ref4} presented evidence that the expansion of the universe
is accelerating. These teams have measured the distances to cosmological supernovae by using the fact that the intrinsic luminosity of Type Ia supernovae
is closely correlated to their decline rate from maximum brightness, which can be independently measured. These measurements, combined with red-shift data
for the supernovae, led to the prediction of an accelerating universe. Both team obtained $\Omega_{m}\approx 0.3$ and $\Omega_{\Lambda}\approx 0.7$
and strongly ruled out the traditional $(\Omega_{m}, \Omega_{\Lambda})$= (1, 0) universe. This value of the density parameter
$\Omega_{\Lambda}$ corresponds to a cosmological constant that is small, nevertheless, nonzero and positive.\\
Cosmological (or vacuum energy) constant \cite{ref5}-\cite{ref8} is one of the most theoretical candidate for dark energy. Unfortunately there is a huge difference of order $10^{120}$ between observational ($\Lambda\sim 10^{-55}cm^{-2}$) and the particle physics prediction value for $\Lambda$. This discrepancy is known as cosmological constant problem. Carmeli and Kuzmenko \cite{ref9} have Recently shown that the cosmological relativistic theory \cite{ref10} predicts $\Lambda=1.934\times 10^{-35} s^{-2}$ which is in agreement with the measurements recently obtained by the High-z Supernova Team and Supernova Cosmological Project \cite{ref1}-\cite{ref4}.
There have been Several ans\"{a}tz suggested in which the $\Lambda$ term decays with time \cite{ref11}-\cite{ref25}. Chen and Wu \cite{ref21} have suggested The special ans\"{a}tz  $\Lambda\propto R^{-2}$ (where $R$ is the scale factor of the Robertson-Walker- metric) which has been modified by several authors \cite{ref26}-\cite{ref31}. However, an accelerating universe can not be predicted by all vacuum decaying cosmological models.
Several authors have argued in favor of the dependence $\Lambda\varpropto \frac{1}{t^{2}}$ in different context \cite{ref13}-\cite{ref15}. The relation $\Lambda\varpropto \frac{1}{t^{2}}$ seems to play a major role in cosmology \cite{ref15}. Recently Pradhan et al \cite{ref32} have obtained some LRS Bianchi type II bulk viscous fluid universe with decaying vacuum energy density.\\
A convenient method to describe models close to $\Lambda$CDM is based on the cosmic jerk parameter $j$, a dimensionless third derivative of the scale factor with respect to the cosmic time \cite{ref33}, \cite{ref34}. A deceleration-to-acceleration transition occurs for models with a positive value of $j_{0}$ and negative $q_{0}$. Flat $\Lambda$CDM models have a constant jerk $j = 1$.\\

Stiff fluid cosmological models create more interest in the study because for these models, the speed of light is equal to speed of
sound and its governing equations have the same characteristics as those of gravitational field (Zel'dovich \cite{ref35}). Barrow \cite{ref36} has discussed
the relevance of stiff equation of state $\rho=p$ to the matter content of the universe in the early state of evolution of universe. Wesson \cite{ref37}
has investigated an exact solution of Einstein's field equation with stiff equation of state. Mohanty et al. \cite{ref38} have investigated cylindrically
symmetric Zel'dovich fluid distribution in General Relativity. G\"{o}tz \cite{ref39} obtained a plane symmetric solution of Einstein's field equation for
stiff perfect fluid distribution. Bali and Tyagi \cite{ref40} have investigated Bianchi type I magnetized stiff fluid cosmological model in General
Relativity.\\

In this latter, a new anisotropic L.R.S. (Locally Rotationally Symmetric) Bianchi type II stiff fluid cosmological model with variable
$\Lambda$ has been investigated by assuming a supplementary condition $A=B^{m}$ between metric potentials $A,~B$ where $n$ is the constant. The out line of the paper is as follows: In Section 2, the metric and the field equations are described. Section 3 deals with the solutions of the field equations. In Subsections (3.1) some physical and
geometric properties of the model are described. Section 4 the jerk parameter of this model is driven. Finally, conclusions are summarized in the last Section 5.
\section{The Metric and Field Equations}
The metric for LRS Bianchi type II in an orthogonal frame is given by
\begin{equation}
\label{eq1} ds^{2}=g_{ij}\theta^{i}\theta^{j},\qquad g_{ij}=diag(-1,1,1,1)
\end{equation}
where the Cartan bases $\theta^{i}$ are given by
\begin{equation}
\label{eq2} \theta^{0}=dt,\qquad \theta^{1}=B\omega^{1},\qquad \theta^{2}=A\omega^{2},\qquad \theta^{3}=A\omega^{3}
\end{equation}
Here, $A$ and $B$ are the time-dependent metric functions. Assuming $(x, y, z)$ as local coordinates, the differential one forms $\omega^{i}$ are given by
\begin{equation}
\label{eq3} \omega^{1}=dy+xdz,\qquad \omega^{2}=dz,\qquad \omega^{3}=dx
\end{equation}
The Einstein's cosmological field equations are given by (with $8\pi G=1$ and $c=1$)
\begin{equation}
\label{eq4} R_{ij}-\frac{1}{2}Rg_{ij}+\Lambda g_{ij}=-T_{ij}
\end{equation}
We consider the energy-momentum tensor in the form
\begin{equation}
\label{eq5} T_{ij}=(p+\rho)u_{i}u_{j}+pg_{ij}
\end{equation}
Hence, for energy-momentum tensor and LRS Bianci type II The Einstein's field equations (\ref{eq4}) leads to the following system of equations:
\begin{equation}
\label{eq6} 2\frac{\ddot{A}}{A}+\frac{\dot{A}^{2}}{A^{2}}-\frac{3}{4}\frac{B^{2}}{A^{4}}=-p+\Lambda
\end{equation}
\begin{equation}
\label{eq7} \frac{\ddot{A}}{A}+\frac{\ddot{B}}{B}+\frac{\dot{A}\dot{B}}{AB}+\frac{1}{4}\frac{B^{2}}{A^{4}}=-p+\Lambda
\end{equation}
\begin{equation}
\label{eq8}2\frac{\dot{A}\dot{B}}{AB}+\frac{\dot{A}^{2}}{A^{2}}-\frac{1}{4}\frac{B^{2}}{A^{4}}=\rho+\Lambda
\end{equation}
where an overdot stands for the first and double overdot for second derivative with respect to $t$.\\
The spatial volume for LRS B-II is given by
\begin{equation}
\label{eq9} V= A^{2}B.
\end{equation}
We define $a = (A^{2}B)^{\frac{1}{3}}$ as the average scale factor of LRS B-II model (\ref{eq1})
so that the Hubble's parameter is given by
\begin{equation}
\label{eq10} H = \frac{\dot{a}}{a} = \frac{1}{3}\left(\frac{2\dot{A}}{A} + \frac{\dot{B}}{B}\right).
\end{equation}
We define the generalized mean Hubble's parameter H as
\begin{equation}
\label{eq11} H = \frac{1}{3}(H_{x} + H_{y} + H_{z}),
\end{equation}
where $H_{x} = \frac{\dot{B}}{B}$, $H_{y} = H_{z}=\frac{\dot{A}}{A}$ are
the directional Hubble's parameters in the directions of $x$, $y$ and $z$ respectively. \\\\
The deceleration parameter $q$ is conventionally defined by
\begin{equation}
\label{eq12} q = - \frac{a\ddot{a}}{\dot{a}^{2}}.
\end{equation}
The scalar expansion $\theta$, shear scalar $\sigma^{2}$ and the average anisotropy parameter $Am$
are defined by
\begin{equation}
\label{eq13}
\theta = \frac{2\dot{A}}{A} + \frac{\dot{B}}{B},
\end{equation}
\begin{equation}
\label{eq14}
\sigma^{2} = \frac{1}{2}\left(\sum_{i=1}^{3}H_{i}^{2}-\frac{1}{3}\theta^{2}\right),
\end{equation}
\begin{equation}
\label{eq15} Am = \frac{1}{3}\sum_{i = 1}^{3}{\left(\frac{\triangle
H_{i}}{H}\right)^{2}},
\end{equation}
where $\triangle H_{i} = H_{i} - H (i = 1, 2, 3)$
\section{Solution of the Field Equations}
The field equations (\ref{eq6})-(\ref{eq8}) are a system of three equations with five unknown parameters $A, B, p, \rho, \Lambda$. Two additional constraint relating these parameters are required to obtain explicit solutions of the system. Following  Bali and Jain \cite{ref41} and Pradhan et al. \cite{ref42}, I assume that the expansion ($\theta$) in the model is proportional to the eigen value $\sigma^{1}_{1}$ of the shear tensor $\sigma^{i}_{j}$. This condition leads to
\begin{equation}
\label{eq16} A=B^{m}
\end{equation}
where  $m$ is a constant.\\
In order to overcome the under-determinacy we have here because of
the five un- known involved in three independent field equations,
I assume that the fluid obeys the stiff fluid equation of state
i.e.
\begin{equation}
\label{eq17} p=\rho
\end{equation}
From (\ref{eq6})-(\ref{eq8}), (\ref{eq16}) and (\ref{eq17}) we obtain
\begin{equation}
\label{eq18} 2\ddot{B}+4m\frac{\dot{B}^{2}}{B}=\frac{B^{-4m+3}}{m}+\frac{2\Lambda}{m}B
\end{equation}
Let $\dot{B}=f(B)$ which implies that $\ddot{B}=ff'$, where $f'=\frac{df}{dB}$. Hence (\ref{eq18}) can be written as
\begin{equation}
\label{eq19}\frac{d}{dB}f^{2}+4m\frac{f^{2}}{B}=\frac{B^{-4m+3}}{m}+\frac{2\Lambda}{m}B
\end{equation}
After integrating eq. (\ref{eq19}) leads to
\begin{equation}
\label{eq20}f^{2}=\frac{B^{-4(m-1)}}{4m}+\frac{\Lambda}{m(1+2m)}B^{2}+kB^{-4m}
\end{equation}
where $k$ is an integrating constant.\\
To get deterministic solution in terms of cosmic time $t$ , we suppose $k = 0$. In this case (\ref{eq20}) takes the form
\begin{equation}
\label{eq21} \frac{dB}{\sqrt{\frac{B^{-4(m-1)}}{4m}+\frac{\Lambda}{m(1+2m)}B^{2}}}=dt
\end{equation}
To get deterministic solution, we assume $m=\frac{3}{2}$. In this case eq. (\ref{eq21}), reduces to
\begin{equation}
\label{eq22}\frac{dB}{\sqrt{\frac{1}{6}B^{-2}+\frac{\Lambda}{6} B^{2}}}=dt
\end{equation}
Integrating eq. (\ref{eq22}) we obtain
\begin{equation}
\label{eq23}B^{2}=\sqrt{\frac{1}{\Lambda}}sinh(\sqrt{\frac{2\Lambda}{3}}t)
\end{equation}
and
\begin{equation}
\label{eq24}A^{2}=(\frac{1}{\Lambda})^{\frac{3}{4}}sinh^{\frac{3}{2}}(\sqrt{\frac{2\Lambda}{3}}t)
\end{equation}
Eqs. (\ref{eq23}) and (\ref{eq24}) show that $\Lambda>0$. In this case the LRS Bianchi type II space-time can be written as
\begin{equation}
\label{eq25} ds^{2}=-dt^{2}+\sqrt{\frac{1}{\Lambda}}sinh(\sqrt{\frac{2\Lambda}{3}}t)(dy+xdz)^{2}+(\frac{1}{\Lambda})^{\frac{3}{4}}sinh^{\frac{3}{2}}(\sqrt{\frac{2\Lambda}{3}}t)(dx+dz)^{2}
\end{equation}
\subsection{ The Geometric and Physical Significance of Model}
The energy density $(\rho)$, the pressure $p$ and the vacuum energy
 density $(\Lambda)$ for the model (\ref{eq25}) are given by
\begin{equation}
\label{eq26}p=\rho=\frac{15}{24}\Lambda coth^{2}(\sqrt{\frac{2\Lambda}{3}}t)-\frac{3\Lambda}{4},
\end{equation}
\begin{figure}[htbp]
\centering
\includegraphics[width=8cm,height=8cm,angle=0]{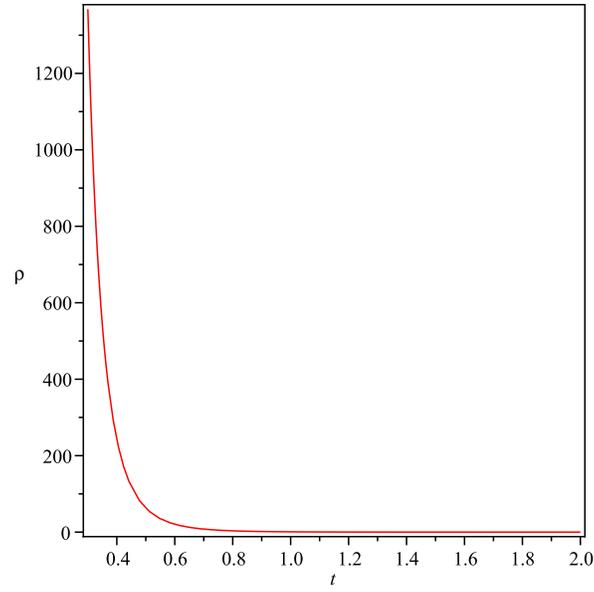}
\caption{The plot of energy density $\rho$ Vs. $t$}
\end{figure}
\begin{figure}[htbp]
\centering
\includegraphics[width=8cm,height=8cm,angle=0]{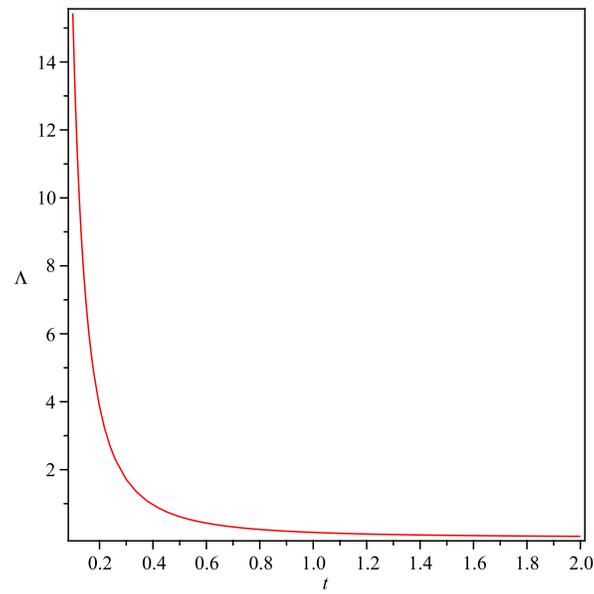}
\caption{The plot of vacuum energy density $\Lambda$ Vs.$t$}
\end{figure}
\begin{equation}
\label{eq27}\Lambda=\frac{3}{2}\left(coth^{-1}\sqrt{\frac{20}{13}}\right)^{2}\frac{1}{t^{2}}.
\end{equation}
From (\ref{eq26}), we see that energy conditions, $\rho\geq 0$ is satisfied under condition
\begin{equation}
\label{eq28} coth^{2}(\sqrt{\frac{2\Lambda}{3}}t)\geq \frac{6}{5}
\end{equation}
From Eq. (\ref{eq26}), it is noted that the proper energy density $\rho(t)$ is a decreasing
function of time and it approaches a small positive value at present epoch. This behavior is clearly depicted in Figures 1.\\
From Eq.(\ref{eq27}), we observe that the cosmological term $\Lambda$ is a decreasing function of time and it approaches
a small positive value at late time. From Figure 2, we note this behavior of cosmological term $\Lambda$
in the model. Recent cosmological observations  suggest the existence of a positive cosmological constant
$\Lambda$ with the magnitude $\Lambda(G\hbar/c^{3})\approx 10^{-123}$. These observations on magnitude and
red-shift of type Ia supernova suggest that our universe may be an accelerating one with induced cosmological
density through the cosmological $\Lambda$-term. Thus, our model is consistent with the results of recent observations. \\

The expressions for Hubble parameter $H$, the scalar of expansion $\theta$, magnitude of shear $\sigma^{2}$, the average anisotropy parameter $A_{m}$  and
the proper volume $V$ for the model (\ref{eq25}) are given by
\begin{equation}
\label{eq29}H=(\frac{2m+1}{3})H_{x}=\frac{2}{3}\sqrt{\frac{2\Lambda}{3}}coth(\sqrt{\frac{2\Lambda}{3}}t),
\end{equation}
\begin{equation}
\label{eq30}\theta=\sqrt{\frac{8}{3}\Lambda}coth(\sqrt{\frac{2\Lambda}{3}}t),
\end{equation}
\begin{equation}
\label{eq31}\sigma^{2}=\frac{2\Lambda}{9}coth^{2}(\sqrt{\frac{2\Lambda}{3}}t),
\end{equation}
\begin{equation}
\label{eq32}A_{m}=2(\frac{m-1}{2m+1})^{2}=\frac{1}{32},
\end{equation}
\begin{equation}
\label{eq33}V=\frac{1}{\Lambda}sinh^{2}(\sqrt{\frac{2\Lambda}{3}}t).
\end{equation}
From (30) and (31) we get
\begin{equation}
\label{eq34}\frac{\sigma^{2}}{\theta^{2}}=constant
\end{equation}
The deceleration parameter is given by
\begin{equation}
\label{eq35} q=-\frac{\ddot{a}a}{\dot{a}^{2}}=-\left[\frac{\frac{4\Lambda}{9}-\frac{4\Lambda}{27}coth^{2}(\sqrt{\frac{2\Lambda}{3}}t)}
{\frac{8\Lambda}{27}coth^{2}(\sqrt{\frac{2\Lambda}{3}}t)}\right]
\end{equation}
If we put the value of $\Lambda$ from eq. (\ref{eq27}) we observe that
\begin{equation}
\label{eq36} q\simeq -0.96.
\end{equation}
From eq. (\ref{eq36}) we observe that our model is in accelerating phase and it's behavior is almost same as the de-sitter universe.\\

Using equations (\ref{eq26})-(\ref{eq29}) we can obtain the matter-energy density $\Omega_{m}$ and dark-energy
density $\Omega_{\Lambda}$ as
\begin{equation}
\label{eq37} \Omega_{m}=\frac{9}{8}\left[\frac{5}{8}-\frac{3}{4}\tanh^{2}\left(coth^{-1}(\sqrt{\frac{20}{13}})\right)\right]\simeq0.155,
\end{equation}
and
\begin{equation}
\label{eq38} \Omega_{\Lambda}=\frac{9}{8}\tanh^{2}\left(coth^{-1}(\sqrt{\frac{20}{13}})\right)\simeq0.731.
\end{equation}
From eqs. (\ref{eq37}) and (\ref{eq38}) we observe that the values of matter-energy density $\Omega_{m}$ and dark-energy density $\Omega_{\Lambda}$
are in good agreement with the values obtain from 5-years WMAP observations for $\Lambda$CMD model \cite{ref43}. The compression of these parameters is shown in table. 1.\\
\begin{table}[h]
\caption{The values of $\Omega_{m}$ and $\Omega_{\Lambda}$ obtain from our model and WMAP}
\centering
\begin{tabular}{c c c c}
\hline\hline{\smallskip}
Parameter & Our model & WMAP\\
\hline
$\Omega_{m}$ & 0.155 & 0.279 \\
$\Omega_{\Lambda}$ & 0.731 & 0.726 \\
\hline
\end{tabular}
\label{table:nonlin}
\end{table}
\section{ The jerk Parameter of the Model}
The jerk parameter in cosmology is defined as the dimensionless third derivative of the scale factor with respect
to cosmic time
\begin{equation}
\label{eq39} j(t)=\frac{1}{H^{3}}\frac{\dot{\ddot{a}}}{a}.
\end{equation}
and in terms of the scale factor
to cosmic time
\begin{equation}
\label{eq40} j(t)=\frac{(a^{2}H^{2})^{''}}{2H^{2}}.
\end{equation}
where the `dots' and `primes' denote derivatives with respect to cosmic time and scale factor, respectively.
 The jerk parameter appears in the fourth term of a Taylor expansion of the scale factor around $a_{0}$
\begin{equation}
\label{eq41} \frac{a(t)}{a_{0}}=1+H_{0}(t-t_{0})-\frac{1}{2}q_{0}H_{0}^{2}(t-t_{0})^{2}+\frac{1}{6}j_{0}H_{0}^{3}(t-t_{0})^{3}+O\left[(t-t_{0})^{4}\right],
\end{equation}
where the subscript $0$ shows the present value. One can rewrite eq. (\ref{eq39}) as
\begin{equation}
\label{eq42} j(t)=q+2q^{2}-\frac{\dot{q}}{H}.
\end{equation}
Using eqs.(\ref{eq29}) and (\ref{eq35}) in (\ref{eq42}) we find
\begin{equation}
\label{eq43} j(t)=\frac{3}{2}\frac{sinh(\sqrt{\frac{2\Lambda}{3}}t)
\left[-2+2sinh(\sqrt{\frac{2\Lambda}{3}}t)
cosh^{3}(\sqrt{\frac{2\Lambda}{3}}t)-3sinh(\sqrt{\frac{2\Lambda}{3}}t)\right]}{cosh^{4}(\sqrt{\frac{2\Lambda}{3}}t)}.
\end{equation}
Now, putting the valu of $\Lambda$ from eq. (\ref{eq27}) in eq. (\ref{eq43}) we obtain
\begin{equation}
\label{eq42} j_{0}= 0.88^{+0.21}_{-0.08}.
\end{equation}
This value does not overlap with the value $j=2.16^{+0.81}_{-0.75}$ obtained from the combination of three kinematical data sets: the
gold sample of type Ia supernovae \cite{ref44}, the SNIa data from the SNLS project \cite{ref45}, and the X-ray galaxy cluster distance measurements \cite{ref46}.
However, it is in consistent with two of the three data sets separately: the SNLS SNIa set gives $j=1.32^{+1.37}_{-1.21}$
 and the cluster set gives $j=0.51^{+2.55}_{-2.00}$, and it is the gold sample data that yields larger $j=2.75^{+1.22}_{-1.10}$ \cite{ref46}.

\section{Concluding Remarks}
\label{sec:4}
A new cosmological model based on LRS Bianchi type II cosmological models with decaying vacuum energy is obtained. The model (\ref{eq25}) starts with a big bang at $t=0$. The expansion in the model decreases as time increases. The proper volume of the model increases as time increases. Since $\frac{\sigma}{\theta}$ is constant the model does not approach isotropy. There is a point type singularity in the model at $t=0$ \cite{ref47}. It is shown that $\Lambda\propto \frac{1}{t^{2}}$. Therefor, as $t\to 0$, $\Lambda\to\infty$ and when $t\to\infty$ then $\Lambda\to 0$. In Brans-Dicke theories the relation like equation (\ref{eq27}) can be finds when one supposes variable gravitational and cosmological ``constant" \cite{ref13}, \cite{ref15} and \cite{ref17}. Berman \cite{ref48} also has derived this relation in general relativity. A positive cosmological constant or equivalently the negative deceleration parameter is required to solve the age parameter and density parameter.\\
The values of deceleration parameter $q$, matter-energy density $\Omega_{m}$, dark-energy density $\Omega_{\Lambda}$ and the jerk parameter for this model are found to be in good agreement with the present values of these parameters obtained from observations. It is reasonable to say that a cosmological model is required to explain acceleration in the present universe. Therefor, the theoretical model found in this paper is in agreement with the recent observations.
\section*{acknowledgements}
Author would like to thank the
Islamic Azad university, Mahshahr branch
for providing facility and support where this
work was carried out.

\end{document}